\RequirePackage{fix-cm}
\documentclass[smallextended]{svjour3}       
\smartqed  
\usepackage{graphicx}
\usepackage{amssymb}
\begin{document}

\title{Exact time evolution of the asymmetric Hubbard dimer
}


\author{Shankar Balasubramanian        \and
        J.~K.~Freericks 
}


\institute{Shankar Balasubramanian \at
              Department of Physics, Massachusetts Institute of Technology, 77 Massachusetts Ave., Cambridge, MA 02139, USA  \\
              Tel.: (703)-609-6320\\
              \email{sbalasub@mit.edu}           
           \and
           J.~K.~Freericks \at
            Department of Physics, Georgetown University, 37th and O Sts. NW, Washington, DC 20057, USA
}

\date{Received: date / Accepted: date}

\maketitle

\begin{abstract}
We examine the time evolution of an asymmetric Hubbard dimer, which has a different on-site interaction on the 
two sites. The Hamiltonian has a time-dependent hopping term, which can be employed to describe an electric field (which creates a Hamiltonian with complex matrix elements), or it can describe a modulation of the lattice (which has real matrix elements). By examining the symmetries under spin and pseudospin, we show that the former case involves at most a $3\times 3$ block---it can be mapped onto the time evolution of a time-independent Hamiltonian, so the dynamics can be evaluated analytically and exactly (by solving a nontrivial cubic equation). We also show that the latter case reduces to at most $2\times 2$ blocks, and hence the time evolution for a single Trotter step can be determined exactly, but the time evolution generically requires a Trotter product.
\end{abstract}

\section{Introduction}
\label{intro}
The two-site asymmetric Hubbard model is an interesting paradigm for the many-body problem, because it can be solved exactly. This was recognized soon after the Hubbard model was introduced in the work of Harris and Falicov~\cite{harris_falicov} for the symmetric case in the 1960s. It turns out that this problem also greatly simplifies the time dynamics, even when the Hamiltonian has an explicit time-dependence and is asymmetric. Recent work has renewed interest in this model~\cite{recent_article}. Here we show how to find the analytic time evolution of the system. When the time-dependent coupling arises from a time-varying complex phase to the hopping, a simple unitary transformation maps the problem onto a time-independent one, which has dynamics that can be determined analytically from the energy eigenvalues. When the time-dependent coupling is a time-varying real function for the hopping, then one can find analytic formulae for the Trotter factor of the evolution, but a full Trotter product needs to be calculated to determine the evolution operator, similar to the Landau-Zener problem. These results can be applied to simplified models of molecules, to the modulation spectroscopy of trapped atoms in a dimer lattice, and also to the two-site Wilson chain, employed as a first step in solving the single-impurity Anderson model. The results discussed here have been presented at the Superstripes 2016 conference~\cite{superstripes}.

\section{The Asymmetric Two-Site Hubbard Model}
\label{sec:1}

We begin with the Hamiltonian for the asymmetric two-site Hubbard model, written in a particle-hole symmetric fashion:

\begin{eqnarray}
\mathcal{H}(t) &=& u_c \left(n^c_{\uparrow} - \frac{1}{2}\right)\left(n^c_{\downarrow} - \frac{1}{2}\right) + u_a \left(n^a_{\uparrow} - \frac{1}{2}\right)\left(n^a_{\downarrow} - \frac{1}{2}\right) \nonumber\\
\label{eq: ham}
&+& \sum_{\sigma} \left [ \gamma(t) c_{\sigma}^{\dagger}a^{\phantom\dagger}_{\sigma} + \gamma^*(t) a_{\sigma}^{\dagger}c^{\phantom\dagger}_{\sigma}\right ].
\end{eqnarray}
The system is determined by two sites and we use the symbol $c_\sigma^{\dagger}$ to denote the creation operator for a fermion of spin $\sigma$ on the first site and $a_\sigma^{\dagger}$ for the second site (the operators $c_\sigma^{\phantom\dagger}$ and $a_\sigma^{\phantom\dagger}$ are the corresponding destruction operators). These operators satisfy the standard anticommutation relations $\{c_\sigma^{\dagger},c_\sigma^{\phantom\dagger}\}_+=1$, 
$\{a_\sigma^{\dagger},a_\sigma^{\phantom\dagger}\}_+=1$, and all other anticommutators between any two fermions vanish. The corresponding number operators are 
$n^c_\sigma=c^{\dagger}_\sigma c^{\phantom\dagger}_\sigma$ and
$n^a_\sigma=a^{\dagger}_\sigma a^{\phantom\dagger}_\sigma$.
The Coulomb repulsion on each site is given by $u_a$ and $u_c$.  The third term is the kinetic energy, where the hopping $\gamma(t)$ is
time-dependent and can be complex.  Because this Hamiltonian has two fermionic degrees of freedom, it can be represented by a $16\times 16$ matrix at each time $t$.  The matrix can be block-diagonalized using symmetries of the system.

\begin{table}
\begin{minipage}{0.65\textwidth}
\caption{Eigenvalues of the 16 different fermionic
states with respect to the conserved symmetries
of the model.}
\label{tab:1}       
\begin{tabular}{llllll}
\hline\noalign{\smallskip}
States & $N$ & $S$ & $S_z$ & $J$ & $J_z$ \\
\noalign{\smallskip}\hline\noalign{\smallskip}
$|0,0\rangle$ & 0 & 0 & 0 & 1 & -1 \\
$|\uparrow,0\rangle$ & 1 & 1/2 & 1/2 & 1/2 & -1/2 \\
$|0,\uparrow\rangle$ & 1 & 1/2 & 1/2 & 1/2 & -1/2 \\
$|\downarrow,0\rangle$ & 1 & 1/2 & -1/2 & 1/2 & -1/2 \\
$|0,\downarrow\rangle$ & 1 & 1/2 & -1/2 & 1/2 & -1/2 \\
$|\uparrow,\uparrow\rangle$ & 2 & 1 & 1 & 0 & 0 \\
$\frac{1}{\sqrt{2}}\left(|\uparrow,\downarrow\rangle + |\downarrow,\uparrow\rangle\right)$ & 2 & 1 & 0 & 0 & 0\\
$\frac{1}{\sqrt{2}}\left(|\uparrow,\downarrow\rangle - |\downarrow,\uparrow\rangle\right)$ & 2 & 0 & 0 & 0 & 0 \\
$\frac{1}{\sqrt{2}}\left(|\uparrow\downarrow, 0\rangle + |0, \uparrow\downarrow\rangle\right)$ & 2 & 0 & 0 & 0 & 0  \\
$\frac{1}{\sqrt{2}}\left(|\uparrow\downarrow, 0\rangle - |0, \uparrow\downarrow\rangle\right)$ & 2 & 0 & 0 & 1 & 0  \\
$|\downarrow,\downarrow\rangle$ & 2 & 1 & -1 & 0 & 0 \\
$|\uparrow\downarrow,\downarrow\rangle$ & 3 & 1/2 & -1/2 & 1/2 & 1/2 \\
$|\downarrow,\uparrow\downarrow\rangle$ & 3 & 1/2 & -1/2 & 1/2 & 1/2 \\
$|\uparrow\downarrow,\uparrow\rangle$ & 3 & 1/2 & 1/2 & 1/2 & 1/2 \\
$|\uparrow,\uparrow\downarrow\rangle$ & 3 & 1/2 & 1/2 & 1/2 & 1/2 \\
$|\uparrow\downarrow,\uparrow\downarrow\rangle$ & 4 & 0 & 0 & 1 & 1 \\
\noalign{\smallskip}\hline
\end{tabular}
\end{minipage}
\begin{minipage}{0.34\textwidth}
\caption{States coupled together by the
hopping term of the Hamiltonian.}
\label{tab:2}       
\begin{tabular}{l}
\hline\noalign{\smallskip}
Couplings  \\
\noalign{\smallskip}\hline\noalign{\smallskip}
$|0,0\rangle$ \\
$|\uparrow, \uparrow\rangle$ \\
$|\downarrow,\downarrow\rangle$ \\
$|\uparrow\downarrow,\uparrow\downarrow\rangle$ \\
$|\uparrow, 0\rangle \to |0, \uparrow\rangle \to |\uparrow, 0\rangle$ \\
$|\downarrow, 0\rangle \to |0, \downarrow\rangle \to |\downarrow, 0\rangle$ \\
$|\uparrow\downarrow, \uparrow\rangle \to |\uparrow, \uparrow\downarrow\rangle \to |\uparrow\downarrow, \uparrow\rangle$ \\
$|\uparrow\downarrow, \downarrow\rangle \to |\downarrow, \uparrow\downarrow\rangle \to |\uparrow\downarrow, \downarrow\rangle$ \\
$|\uparrow, \downarrow\rangle \to |\uparrow\downarrow, 0\rangle \to |\downarrow, \uparrow\rangle \to$\\
$\quad\quad\quad|0, \uparrow\downarrow\rangle \to |\uparrow, \downarrow\rangle$  \\
\noalign{\smallskip}\hline
\end{tabular}
\end{minipage}
\end{table}

The Hubbard model on a bipartite lattice has two $SU(2)$ symmetries: spin and pseudospin. We employ these symmetries to organize the different basis states for the Hamiltonian. This then naturally block diagonalizes $\mathcal{H}(t)$.  To begin, the total number of electrons operator is $N =  n^c_{\uparrow} + n^c_{\downarrow} + n^a_{\uparrow} + n^a_{\downarrow}$, and it commutes with $\mathcal{H}(t)$.  The eigenstates of $N$ and corresponding eigenvalues are indicated in Table \ref{tab:1}.  In this table, we write states in a direct product space $\left\{|0\rangle, |\uparrow\rangle, |\downarrow\rangle, |\uparrow \downarrow\rangle \right\}_c \otimes \left\{|0\rangle, |\uparrow\rangle, |\downarrow\rangle, |\uparrow \downarrow\rangle \right\}_a$.  

The total spin in the z-direction is given by $S_z = \frac{1}{2}\left(n^c_{\uparrow} + n^a_{\uparrow} - n^c_{\downarrow} - n^a_{\downarrow}\right)$, which also commutes with the Hamiltonian.  Similarly, the spin raising and lowering operators, $S_{+} = c_{\uparrow}^{\dagger}c_{\downarrow} + a_{\uparrow}^{\dagger}a_{\downarrow}$ and $S_{-} = c_{\downarrow}^{\dagger}c_{\uparrow} + a_{\downarrow}^{\dagger}a_{\uparrow}$ commute with $\mathcal{H}(t)$.  Hence $S^2=(S_+S_-+S_-S_+)/2+S_zS_z$ and $S_z$ are also good quantum numbers.
The pseudospin operator $J_z = \frac{N}{2} - 1$ always commutes with $\mathcal{H}(t)$ as well. The corresponding raising and lowering operators $J_{+} = c_{\uparrow}^{\dagger}c_{\downarrow}^{\dagger} - a_{\uparrow}^{\dagger}a_{\downarrow}^{\dagger}$ and $J_{-} = c_{\downarrow}c_{\uparrow} - a_{\downarrow}a_{\uparrow}$ commute only when $\gamma(t)$ is real, so in general, $J^2=(J_+J_-+J_-J_+)/2+J_zJ_z$ is not a conserved symmetry.  The quantum numbers under all of these symmetries are summarized in Table~\ref{tab:1}. Using these symmetries, we immediately see that the matrix should block diagonalize into four $1\times 1$ blocks, four $2\times 2$ blocks, and, if $J$ is a good quantum number, two more $1\times 1$ and one more $2\times 2$ block, otherwise one more $1\times 1$ and one $3\times 3$ block.

Table \ref{tab:2} shows which states are coupled together by the hopping term. In this case, we use the direct product states to show the couplings, and in this basis, the Hamiltonian block diagonalizes to four $1\times 1$ blocks, four $2\times 2$ blocks and a $4\times 4$ block.  We enumerate the states in the following order: ${|0,0\rangle},\allowbreak {|\uparrow,0\rangle},\allowbreak {|0,\uparrow\rangle},\allowbreak {|\downarrow,0\rangle},\allowbreak {|0,\downarrow\rangle},\allowbreak {|\uparrow,\uparrow\rangle},\allowbreak {|\uparrow,\downarrow\rangle},\allowbreak {|\uparrow\downarrow,0\rangle},\allowbreak {|0,\uparrow\downarrow\rangle},\allowbreak {|\downarrow,\uparrow\rangle},\allowbreak
{|\downarrow,\downarrow\rangle},\allowbreak {|\uparrow\downarrow,\downarrow\rangle},\allowbreak {|\downarrow,\uparrow\downarrow\rangle},\allowbreak  {|\uparrow\downarrow,\uparrow\rangle},\allowbreak {|\uparrow,\uparrow\downarrow\rangle},\allowbreak {|\uparrow\downarrow,\uparrow\downarrow\rangle}$.  This labels the states in order of total number operator eigenstates, and pairs together states which are coupled through the Hubbard Hamiltonian.  The $1\times 1$ blocks are summarized in Table \ref{tab:3}; the remaining blocks are given next.  The $2\times 2$ blocks for the $|\uparrow,0\rangle$ and $|0,\uparrow\rangle$ states and for the $|\downarrow,0\rangle$ $|0,\downarrow\rangle$ states are identical and given by
\begin{equation}
\mathcal{H}^{(1)}_{2\times 2}(t) = \left(
\begin{array}{cc}
 -\frac{1}{4}\left(u_c - u_a\right) & \gamma^*(t) \\
 \gamma(t) & \frac{1}{4}\left(u_c - u_a\right) \\
\end{array}
\right).
\end{equation}
For the $|\uparrow\downarrow,\downarrow\rangle$ and $|\downarrow,\uparrow\downarrow\rangle$ states and the $|\uparrow\downarrow,\uparrow\rangle$ and $|\uparrow,\uparrow\downarrow\rangle$ states the $2\times 2$ blocks are also identical and given by
$\mathcal{H}^{(2)}_{2\times 2}(t) = -\mathcal{H}^{(1)}_{2\times 2}(t)$.
The remaining $4 \times 4$ block, for the four states $\left\{|\uparrow,\downarrow\rangle, |\uparrow\downarrow,0\rangle, |0,\uparrow\downarrow\rangle, |\downarrow,\uparrow\rangle\right\}$ becomes
\begin{equation}
\mathcal{H}_{4\times 4} = \left(
\begin{array}{cccc}
 -\frac{1}{4}\left(u_c + u_a\right) & \gamma(t) & \gamma^*(t) & 0 \\
 \gamma^*(t) & \frac{1}{4}\left(u_c + u_a\right) & 0 & -\gamma^*(t) \\
 \gamma(t) & 0 & \frac{1}{4}\left(u_c + u_a\right) & -\gamma(t) \\
 0 & -\gamma(t) & -\gamma^*(t) & -\frac{1}{4}\left(u_c + u_a\right) \\
\end{array}
\right).
\end{equation}
To simplify this further, we employ the $S$ and $J$ symmetries.  Converting to the basis given in Table \ref{tab:1} yields
\begin{equation}
\mathcal{H}_{4\times 4}(t) = \left(
\begin{array}{cccc}
 -\frac{1}{4}\left(u_c + u_a\right) & 0 & 0 & 0 \\
 0 & -\frac{1}{4}\left(u_c + u_a\right) & \gamma(t) + \gamma^*(t) & \gamma(t) - \gamma^*(t) \\
 0 & \gamma(t) + \gamma^*(t) & \frac{1}{4}\left(u_c + u_a\right) & 0 \\
 0 & \gamma^*(t) - \gamma(t) & 0 & \frac{1}{4}\left(u_c + u_a\right) \\
\end{array}
\right).
\end{equation}
When $\gamma(t) = \gamma^*(t)$, we see that the Hamiltonian decomposes into a $2\times 2$ and two $1\times 1$ blocks, as claimed because $J$ is a good quantum number.

\begin{table}
\caption{$1\times 1$ blocks of $\mathcal{H}(t)$ and their corresponding $1\times 1$ blocks in the evolution operator.}
\label{tab:3}       
\begin{tabular}{lll}
\hline\noalign{\smallskip}
State & Hamiltonian Matrix Element & Evolution Operator Matrix Element \\ 
\noalign{\smallskip}\hline\noalign{\smallskip}
$|0,0\rangle$ & $~~\left(u_c + u_a\right)/4$ & $e^{-i \frac{t}{4}\left(u_c + u_a\right)}$\\
$|\uparrow, \uparrow\rangle$ & $-\left(u_c + u_a\right)/4$ & $e^{i \frac{t}{4}\left(u_c + u_a\right)}$\\
$|\downarrow,\downarrow\rangle$ & $-\left(u_c + u_a\right)/4$ & $e^{i \frac{t}{4}\left(u_c + u_a\right)}$\\
$|\uparrow\downarrow,\uparrow\downarrow\rangle$ & $~~\left(u_c + u_a\right)/4$ & $e^{-i \frac{t}{4}\left(u_c + u_a\right)}$\\
\noalign{\smallskip}\hline
\end{tabular}
\end{table}

 Now, we move on to describe the evolution operator, which is given by
 \begin{equation}
 \mathcal{U}(t,t')=\mathcal{T}_t e^{-i\int_{t'}^t d\bar t \mathcal{H}(\bar t)}=\mathcal{U}(t,t-\Delta t)\mathcal{U}(t-\Delta t,t-2\Delta t)\cdots\mathcal{U}(t'+\Delta t,t'),
 \end{equation}
 where $\mathcal{T}_t$ denotes the time-ordered product and the RHS is a Trotter product.
If we numerically evaluate the evolution operator using a Trotter product, the evolution operator can be determined analytically at each time step. For complex $\gamma(t)$, we must exponentiate a $3\times 3$ matrix, which requires us to solve a cubic equation.  However, for real $\gamma(t)$, the total Hamiltonian is only composed of $2 \times 2$ blocks at worst.  An example of a time-dependent $\gamma(t)$ that is real arises in a modulation spectroscopy experiment, where the hopping is modulated in magnitude via $\gamma(t)=\gamma_0+\delta\gamma\cos(\Omega t)$, although the specific functional form for the real $\gamma(t)$ is not needed to determine the overall dynamics of the system. The evolution operator for the $1\times 1$ blocks are tabulated in Table \ref{tab:3}.  The Trotter factor for the $2 \times 2$ blocks with real $\gamma$ are
\begin{equation}
\mathcal{U}^{(1)}_{2\times 2} = \cos \left(\Delta t\sqrt{\frac{\delta u^2}{16} +  \gamma^2}\right)\mathbb{I} + \frac{ i \sin \left(\Delta t\sqrt{\frac{\delta u^2}{16} + \gamma^2}\right)}{\sqrt{\frac{\delta u^2}{16} +  \gamma^2}}\left(\gamma \tau_x - \frac{\delta u}{4}  \tau_z\right),
\end{equation}
and $\mathcal{U}^{(2)}_{2\times 2}=\mathcal{U}^{(1)}_{2\times 2}|_{t\rightarrow -t}$,
where $\delta u= u_c - u_a$, and $\tau$ represents a Pauli spin matrix.  Note that $\Delta t$ here represents a small time-interval and we suppressed the time dependence of $\gamma$.  For the $4\times 4$ block, the Trotter factor is given by
\begin{equation}
 \mathcal{U}_{4\times 4} = \left(
\begin{array}{c|c|c}
 e^{i\frac{\Delta ts}{4}} & 0 &0  \\
 \hline
  0 &\mathcal{U}^{(3)}_{2\times 2} & 0 \\
  \hline
  0 & 0 & e^{-i\frac{\Delta ts}{4}} \\
\end{array}
\right),
\end{equation}
with
\begin{equation}
\mathcal{U}^{(3)}_{2\times 2} = \cos \left(\Delta t\sqrt{\frac{s^2}{16} + 4 \gamma^2}\right) + \frac{ i \sin \left(\Delta t\sqrt{\frac{s^2}{16} + 4 \gamma^2}\right)}{\sqrt{\frac{s^2}{16} + 4 \gamma^2}}\left(2 \gamma \tau_x - \frac{s}{4}  \tau_z\right),
\end{equation}
where $s = u_c + u_a$.  This allows for the complete solution of the time-evolution using the analytic Trotter factors.

Another common time-dependent problem is when an electric field is applied to the system. The dc field (turned on at $t=0$) is described by a time-dependent vector potential $\mathbf{E}=-d\mathbf{A}(t)/dt$ with $\mathbf{A}(t)=-\mathbf{E}t\theta(t)$ and $\theta(t)$ the unit step function. Then the functional form becomes $\gamma(t) = \gamma_0e^{-i\mathbf{A}(t)\cdot(\mathbf{R}_a-\mathbf{R}_c)}= \gamma_0 e^{i E t}$ if we assume the $a$ site is to the right of the $c$ site and $\gamma_0$ is real.  The $4 \times 4$ block (in the original direct product basis) of the Hamiltonian is converted to $\bar \mathcal{H}_{4\times 4}=U^\dagger\mathcal{H}_{4\times 4}U-iU^{\dagger}\dot U$ via the unitary transformation $U=\mathrm{diag}\{1,\exp(-iEt),\exp(iEt),1\}$ which yields
\begin{equation}
\tilde{\mathcal{H}}_{4\times 4} = \left(
\begin{array}{cccc}
 -\frac{1}{4}\left(u_c + u_a\right) & \gamma_0 & \gamma_0 & 0 \\
 \gamma_0 & \frac{1}{4}\left(u_c + u_a\right) - E & 0 & -\gamma_0 \\
 \gamma_0 & 0 & \frac{1}{4}\left(u_c + u_a\right) + E & -\gamma_0 \\
 0 & -\gamma_0 & -\gamma_0 & -\frac{1}{4}\left(u_c + u_a\right) \\
\end{array}
\right).
\end{equation}
This is now time-independent, so the evolution operator can be directly written down.  Converting to the $S$ and $J$ eigenstate basis, we find
\begin{equation} \label{eq:1}
\tilde{\mathcal{H}}_{4\times 4} = \left(
\begin{array}{cccc}
 -\frac{1}{4}s & 0 & 0 & 0 \\
 0 & \frac{1}{4}s & 2 \gamma_0 & 0 \\
 0 & 2 \gamma_0 & \frac{1}{4}s & -E \\
 0 & 0 & -E & -\frac{1}{4}s \\
\end{array}
\right).
\end{equation}
The evolution operator can now be analytically obtained using the cubic formula and multiplying by the diagonal matrix $U$.

For the other $2 \times 2$ blocks, we similarly move to a rotating frame with $U' = \mathrm{diag}\left\{1, e^{i E t}\right\}$, and we find
\begin{equation}
\mathcal{H}^{(1)}_{2\times 2} = \left(
\begin{array}{cc}
 -\frac{1}{4}\delta u & \gamma_0 \\
 \gamma_0 & E + \frac{1}{4}\delta u \\
\end{array}
\right),
\quad
\mathcal{H}^{(2)}_{2\times 2} = \left(
\begin{array}{cc}
 \frac{1}{4}\delta u & -\gamma_0 \\
 -\gamma_0 & E - \frac{1}{4}\delta u \\
\end{array}
\right).
\end{equation}

Note that the time-independent form of the Hamiltonian corresponds precisely to the case in Ref.~\cite{recent_article}, with $E=-\Delta v$, as expected (when we take the limit $u_c=u_a$).

\section{Results}

For the case of real $\gamma$, we choose $\gamma(t) = \gamma_0 + \delta \gamma \cos \Omega t$; we pick $u_c=2$, $u_a=1$, $\gamma_0=1$ and $\delta\gamma=0.5$.  We measure the  double occupancy, or $\langle n^c_{\uparrow}n^c_{\downarrow}\rangle $ and $\langle n^a_{\uparrow}n^a_{\downarrow}\rangle$.  Modulating 
the lattice for cold atoms typically leads to an increase of double occupancy along with oscillations. We use
\begin{equation}
D_g^\alpha(t) = \langle \psi_g | \mathcal{U}^{\dagger(3)}_{2 \times 2}(t,0) n_{\alpha\uparrow}n_{\alpha\downarrow} \mathcal{U}^{(3)}_{2 \times 2}(t,0)| \psi_g \rangle,
\end{equation}
\begin{equation}
D_t^\alpha(t) = \textrm{Tr}_{N=2}\left(e^{-\beta \mathcal{H}(0)}\mathcal{U}^{\dagger}(t,0) n_{\alpha\uparrow}n_{\alpha\downarrow} \mathcal{U}(t,0)\right),
\end{equation}
where we can restrict the trace to the $N=2$ sector for the canonical thermal distribution and we take $\alpha=a,c$.

\begin{figure}
  \includegraphics[width=1.01\textwidth]{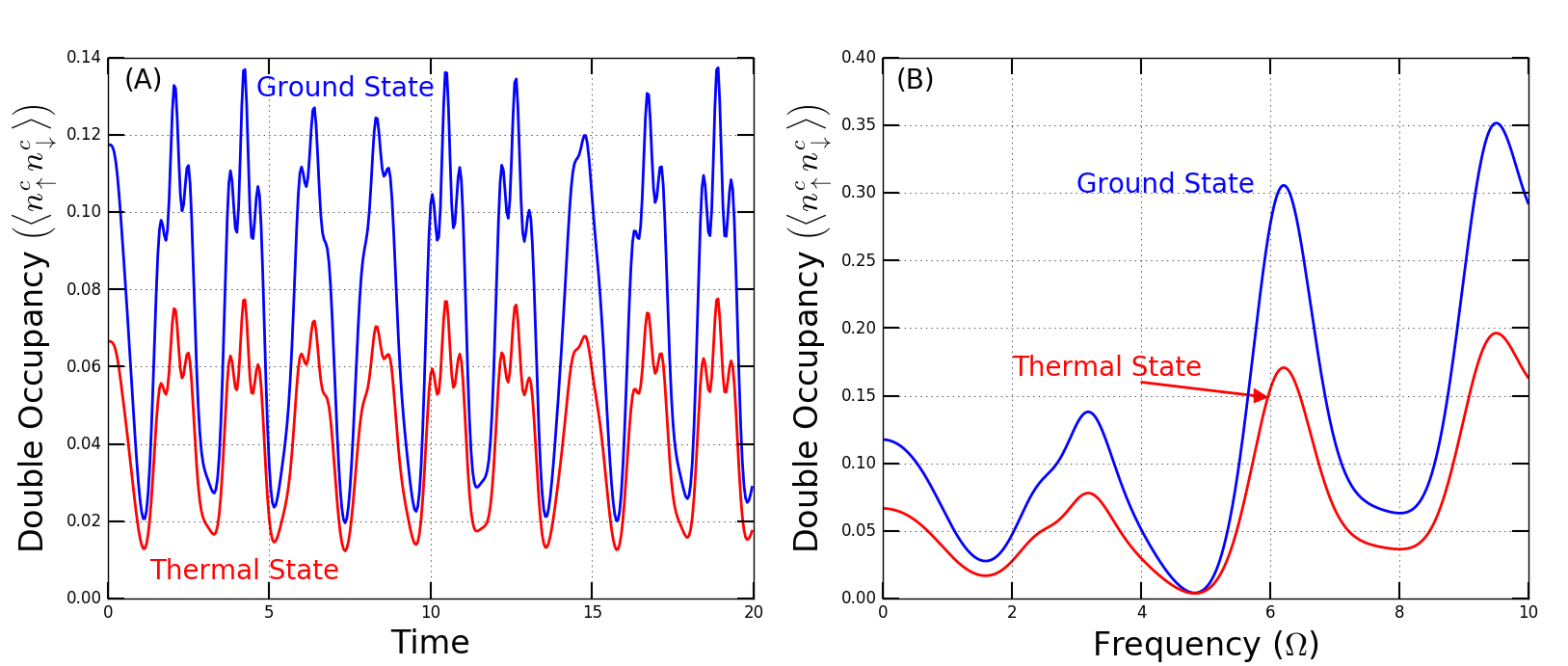}
\caption{Double occupancy for $c$ (A) as a function of time for $\Omega=3$ and (B) as a function of frequency for $t=2$.  The thermal state has $\beta = 0.5$.  The functional form for the modulation is $\gamma(t) = 2 + \cos(\Omega t)$; the interactions are $u_c = 10$ and $u_a = 5$. As expected the periodicity remains the same at finite temperature, but with a reduced amplitude. }
\label{fig:1}       
\end{figure}

Figure \ref{fig:1} shows the double occupancy for a simplified modulation spectroscopy, both as a function of time at fixed frequency and as a function of frequency at fixed time.  As one might expect, the amplitude of the oscillations is smaller for a thermal state than for the ground state. The curves don't show perfect periodicity, because there is more than one frequency contributing to the oscillations. But, the dominant frequency is the driving frequency $\Omega$. In panel (B), we can see that there are broad resonances when $\Omega$ is a multiple of 3.

For the dc electric field, we have $\gamma(t) = \gamma_0 e^{i E t}$.  We measure the polarization of the molecule given by $\langle n^c_{\uparrow} + n^c_{\downarrow} - n^a_{\uparrow} - n^a_{\downarrow}\rangle$.  Similar to the double occupancy, we have
\begin{equation}
P_g(t) = \langle \psi_g | \mathcal{U}^{\dagger}_{4 \times 4}(t,0) (n_c - n_a) \mathcal{U}_{4 \times 4}(t,0)| \psi_g \rangle,
\end{equation}
\begin{equation}
P_t(t) = \textrm{Tr}_{N=2}\left(e^{-\beta \mathcal{H}(0)}\mathcal{U}^{\dagger}(t,0) (n_c - n_a) \mathcal{U}(t,0)\right),
\end{equation}
where we are working with a canonical distribution that fixes $N=2$ for the thermal distribution.  

\begin{figure}
  \includegraphics[width=1.01\textwidth]{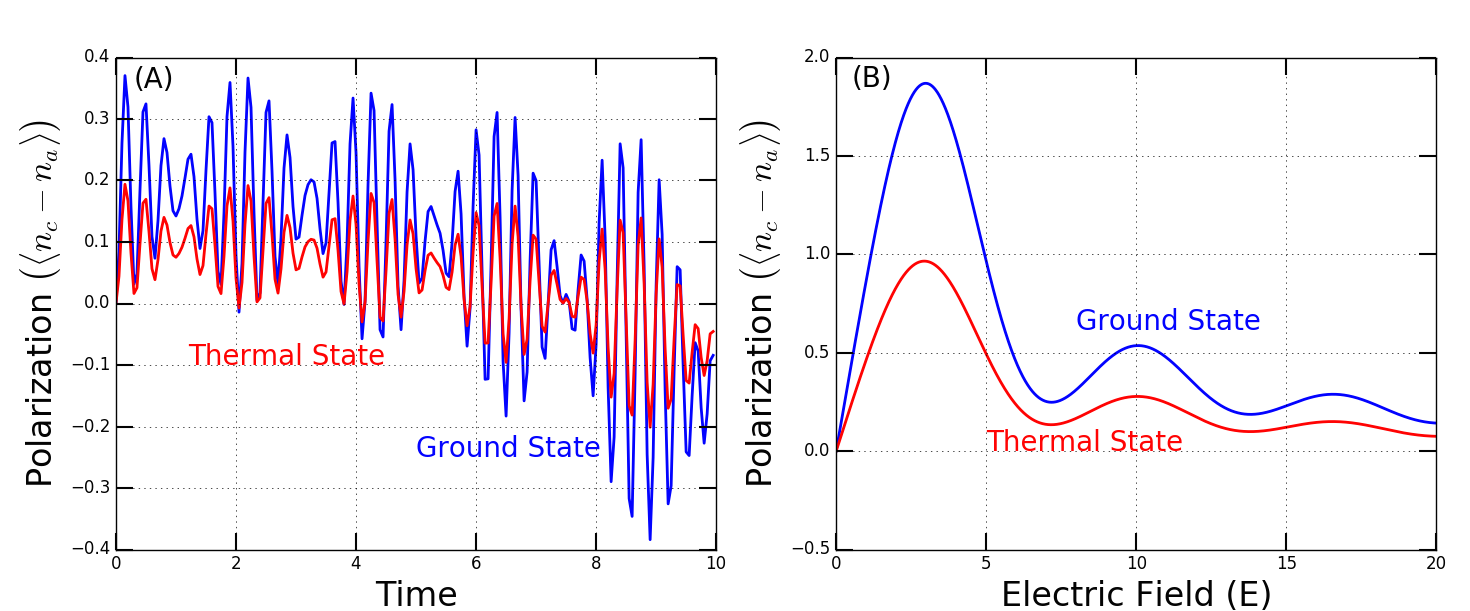}
\caption{Polarization with respect to time with $E=20$ (A) and electric field with $t=1$ (B).  In both cases, $u_a = 2$, $u_c = 1$ and $\beta = 1$.  The field strength $\gamma_0 = 1$.  }
\label{fig:2}       
\end{figure}

The polarization is shown in Figure \ref{fig:2}, where the blue represents the ground state polarization, and the red represents the thermal polarization with $\beta=1$.  For most cases we observed that the thermal polarizability rapidly converges to the ground state polarizability as one increases $\beta$, and they look identical for $\beta \sim 10$.  
The polarization also can develop beats.  For the ground state polarization, there are two dominant frequencies, and a third less prominent frequency; this comes from the three energies in the $3\times 3$ block and the fact that the electric field does not change the symmetry of the wavefunctions as $t$ varies.
The thermal state is somewhat more complicated, showing more frequencies and a reduced amplitude as expected.
The electric field dependence shows a decreasing amplitude
with increasing $E$ and oscillations at a single frequency for the given time.

\begin{figure}
  \includegraphics[width=1.01\textwidth]{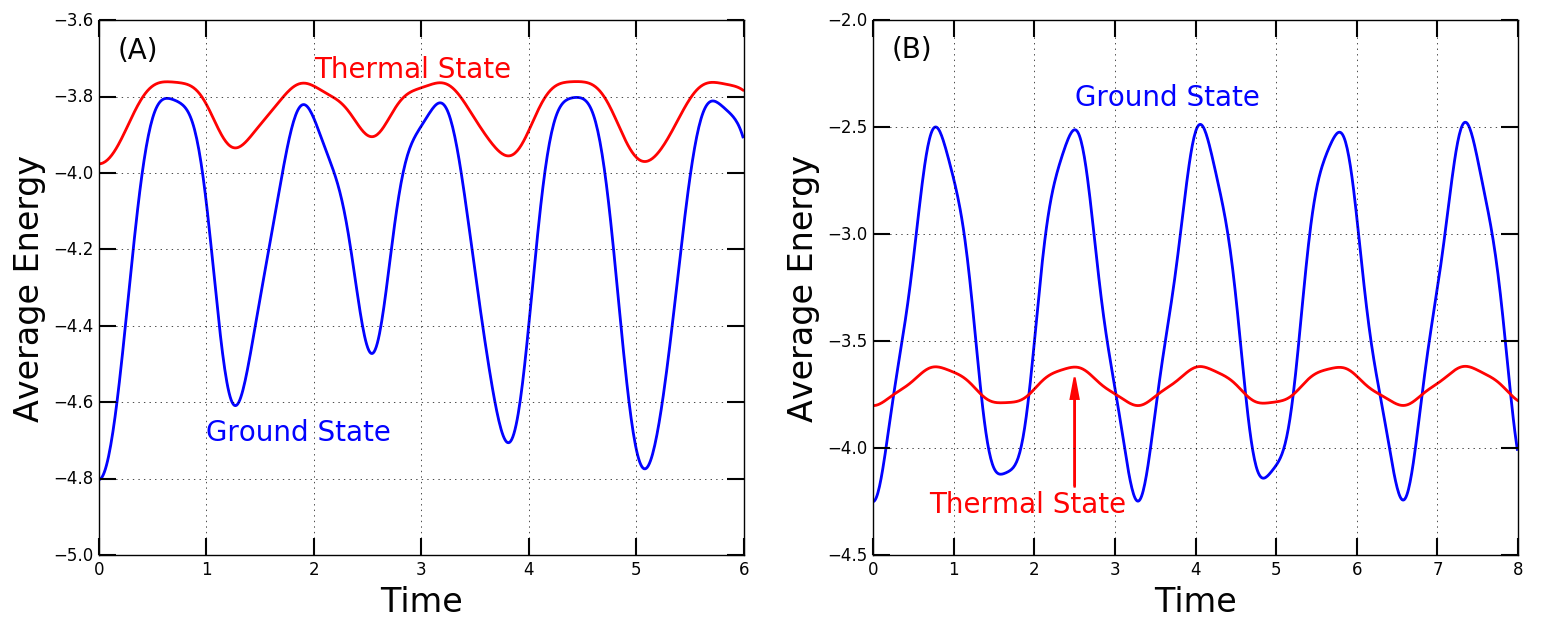}
\caption{Average energy as a function of time for $\gamma(t) = 1 + 0.5 \cos \Omega t$ (A) and $\gamma(t) = e^{i E t}$ (B).  We have $u_c = 10$, $u_a = 5$, $\beta = 1$, and $E = \Omega = 5$.
}
\label{fig:3}       
\end{figure}

In Figure \ref{fig:3}, we show the average energy as a function of time for both cases.  They oscillate approximately at the expected frequencies of $2\pi/\Omega$ and $2\pi/E$. In spite of the fact that this is an interacting system, the energy tends to oscillate and not have any overall increase with time.

While these results, and this model are quite simple, there is experimental evidence of two different sites in the strongly correlated CuO$_2$ plane of cuprates, which has been discussed as a generic  feature of high temperature superconductivity~\cite{bianconi} and provides some experimental support for the features discussed in the theory above.

\section{Conclusions}

We show that the real time behavior of the asymmetric two-site Hubbard model can be mapped to a block-diagonal time-independent 
Hamiltonian when the coupling is given by an electric field, and can be described by the Trotter formula with exact analytic results
for each Trotter factor when the off-diagonal hopping is always real. We used these solutions to examine the double occupancy 
for an analog of modulation spectroscopy and the polarization for an electric field. In future work, we plan to incorporate this time evolution into the solution of quantum impurity problems, which will further broaden their applicability.

\begin{acknowledgements}
This work was supported by the Department of Energy, Office of Basic Energy Sciences, Division of Materials Sciences and Engineering under Contract No. DE-FG02-08ER46542.
S. B. was also supported by the National Science Foundation under grant number PHY-1314295.
J.K.F. was also supported by the McDevitt Bequest at Georgetown. 
\end{acknowledgements}


\begin{thebibliography}{}
%
%
\bibitem{harris_falicov}
L. M. Falicov and R. A. Harris, J. Chem. Phys. {\bf 51}, 3153--3158 (1969). 
\bibitem{recent_article}
J. I. Fuks and N. T. Maitra, Phys. Rev. A {\bf 89}, 062502 (2014).
\bibitem{superstripes}
T. Adachi et al. {\it Superstripes 2016}, edited by A. Bianconi (Superstripes Press, Rome, It, 2016) Isbn: 9788866830559.

\bibitem{bianconi}
A. Bianconi, Int. J. Mod. Phys. B {\bf 14}, 3289-3297 (2000). 

\end{thebibliography}


\end{document}